\documentclass[twocolumn]{aastex7}

\usepackage{amsmath}
\usepackage{booktabs}

\begin{document}

\title{Revisiting the Role of Global Magnetic Variables in Coronal Heating of Solar Active Regions} 

\author[0000-0002-2189-9313]{David H. Brooks}
\affiliation{Computational Physics Inc., Springfield, VA 22151, USA}
\affiliation{University College London, Mullard Space Science Laboratory, Holmbury St. Mary, Dorking, Surrey, RH5 6NT, UK}
\email{dhbrooks.work@gmail.com}

\author[0000-0001-6102-6851]{Harry P. Warren}
\affiliation{Space Science Division, Naval Research Laboratory, Washington, DC 20375, USA}
\email{harry.warren@nrl.navy.mil}

\author[0000-0002-6495-4685]{Johnathan R. Stauffer}
\affiliation{Space Science Division, Naval Research Laboratory, Washington, DC 20375, USA}
\email{johnathan.r.stauffer.ctr@us.navy.mil}

\begin{abstract}
Reconnection-based models of coronal heating predict that magnetic complexity---non-potential
configurations, currents, shear---should enhance coronal emission independently of
total magnetic flux.  Previous work using Yohkoh soft X-ray observations and ground-based vector
magnetograms, however, found no evidence that complexity contributes independently to coronal
heating \citep{Fisher1998}. We revisit this question using SDO/HMI vector magnetograms and SDO/AIA EUV imaging.
HMI Space-weather Active Region Patches (SHARPs) provide 16 magnetic parameters
characterizing field strength, gradients, currents, helicity, and non-potentiality. We measure
spatially-integrated transition region and
coronal emission using AIA 304~\AA, 193~\AA, 94~\AA,
and isolated Fe~\textsc{xviii} emission
for 137 active regions, sampling plasma at
$\sim$0.09~MK, $\sim$1.4~MK, and $\sim$7.1~MK.
Partial correlation analysis isolates the contribution of each
complexity parameter after controlling for total unsigned vertical magnetic flux. Unsigned flux
($\Phi_M$) explains 72--82\% of variance in emission, following power laws ($I
\propto \Phi_M^{\alpha}$, where $\alpha$=0.61-1.19) consistent with prior soft X-ray results. Partial correlations reveal
weak positive contributions ($r = 0.2-0.3$) to transition region (304~\AA) emission from some
vertical current and field parameters,
but little evidence that other active-region-averaged measures of magnetic complexity
enhance coronal emission.
The exception is total photospheric magnetic free energy density
($\rho_{\mathrm{tot}}$), which shows a \emph{positive} partial correlation ($r = 0.49$, $p <
0.001$) with Fe~\textsc{xviii} emission, but
\emph{negative}, or non-existent, partial correlations with other AIA filters.
These results suggest quasi-steady coronal heating
scales with magnetic flux, while $\rho_{\mathrm{tot}}$ may contribute to high temperature coronal heating.
At lower temperatures, regions with more stored energy than expected for their flux produce less EUV emission
than expected.
\end{abstract}


\section{Introduction} 

\label{sec:introduction}

The solar corona is structured and heated by magnetic fields. Energy generated in the convection zone is transported along field lines into the corona, where it is dissipated and radiated away. Understanding what controls this energy input is central to the coronal heating problem.

Observationally, a robust relationship exists between photospheric magnetic flux and coronal emission. Soft X-ray luminosity scales with total unsigned magnetic flux across active regions spanning orders of magnitude in size \citep{Fisher1998,Pevtsov2003,Yoshimura2025}.
This relationship extends to EUV emission \citep{Warren2012}, radio observations, and even stellar coronae \citep{Toriumi2022}. The consistency of this scaling suggests that magnetic flux sets the overall level of coronal heating.

Yet flux alone does not specify how heating occurs. Models based on magnetic reconnection—the nanoflare hypothesis—predict that heating should depend on magnetic complexity \citep{Klimchuk2006}. Non-potential field configurations store free magnetic energy in the form of currents, shear, and twist. When this energy is released through small-scale reconnection events, it heats the coronal plasma. If nanoflares dominate coronal heating, then measures of magnetic complexity should correlate with coronal emission beyond what total flux explains.

\citet{Fisher1998} tested this prediction using Yohkoh soft X-ray images and ground-based vector magnetograms. They found that soft X-ray luminosity scales as a power law with total unsigned vertical magnetic flux ($L_X \propto \Phi^{1.19}$), but found no evidence that magnetic twist or shear contributes additional heating. After accounting for flux, residual correlations with complexity measures were weak and statistically insignificant. Their conclusion was direct: total magnetic flux determines coronal heating, and observable complexity does not.

\begin{figure*}[ht!]
  \centerline{%
    \includegraphics[width=0.8\textwidth]{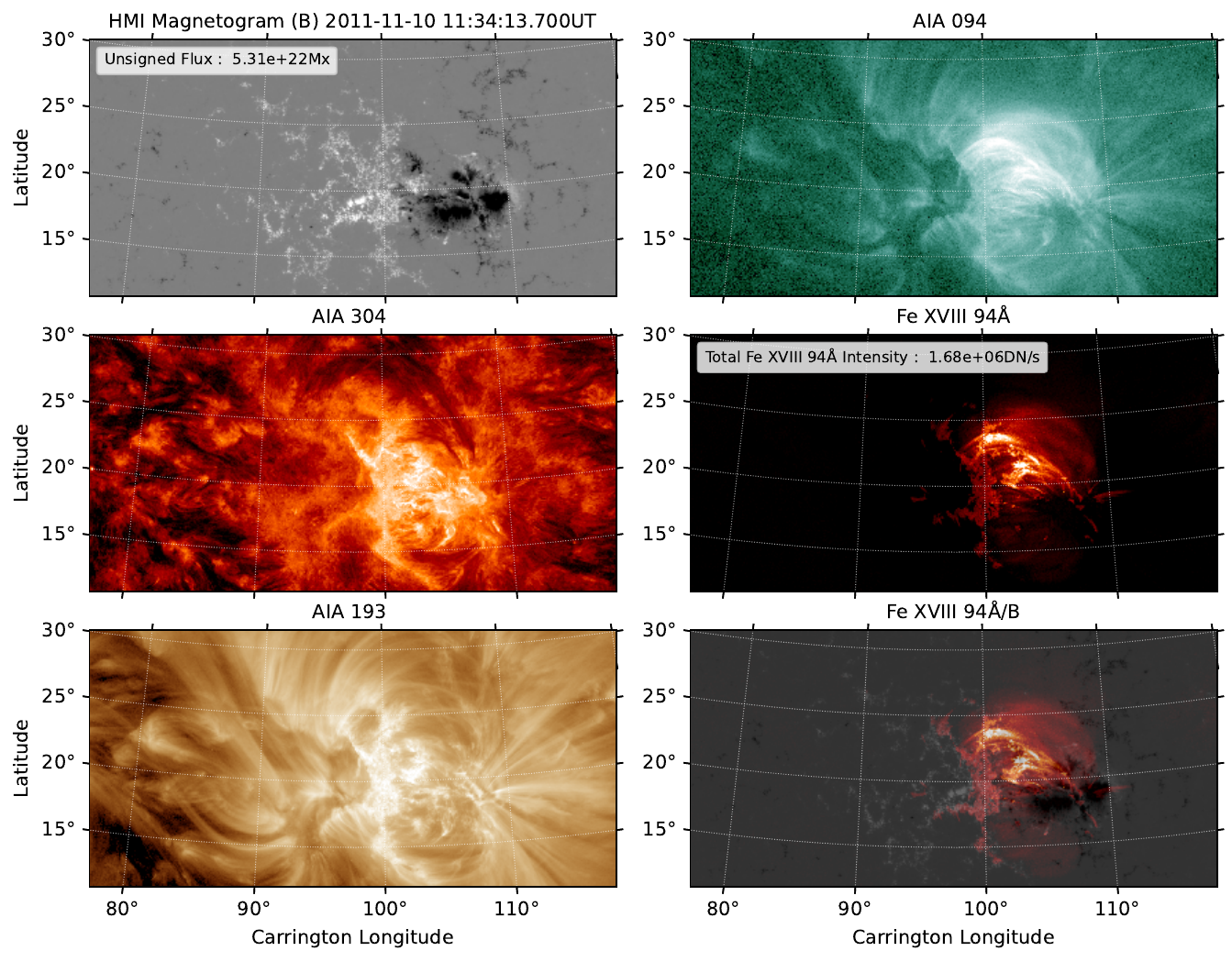}} %
\caption{ Overview of AR 11339 observed on 2011 November 10 and included in our sample. 
Left panels (top): HMI line-of-sight magnetogram extracted from the SHARPs dataset. We show the pre-computed total unsigned magnetic flux from the SHARPs header in the legend. 
Center: AIA 304~\AA\ image showing the lower transition region at $\sim$0.09~MK.
Bottom: AIA 193~\AA\ image showing the corona at $\sim$1.4~MK.
Right panels (top): AIA 94~\AA\ image showing the corona at $\sim$7.1~MK.
Center: Image showing the Fe~\textsc{xviii} 93.926~\AA\ intensity after isolation from the AIA 94~\AA\ channel using the method of \cite{Warren2012}. We show the total intensity in the SHARPs field-of-view (FOV) in the legend.
Bottom: Fe~\textsc{xviii} 93.926~\,\AA\ image overlaid on the HMI magnetogram to show the accuracy of the co-alignment procedure.
\label{fig1}}
\end{figure*}

\begin{figure*}[ht!]
  \centerline{%
    \includegraphics[width=0.8\textwidth]{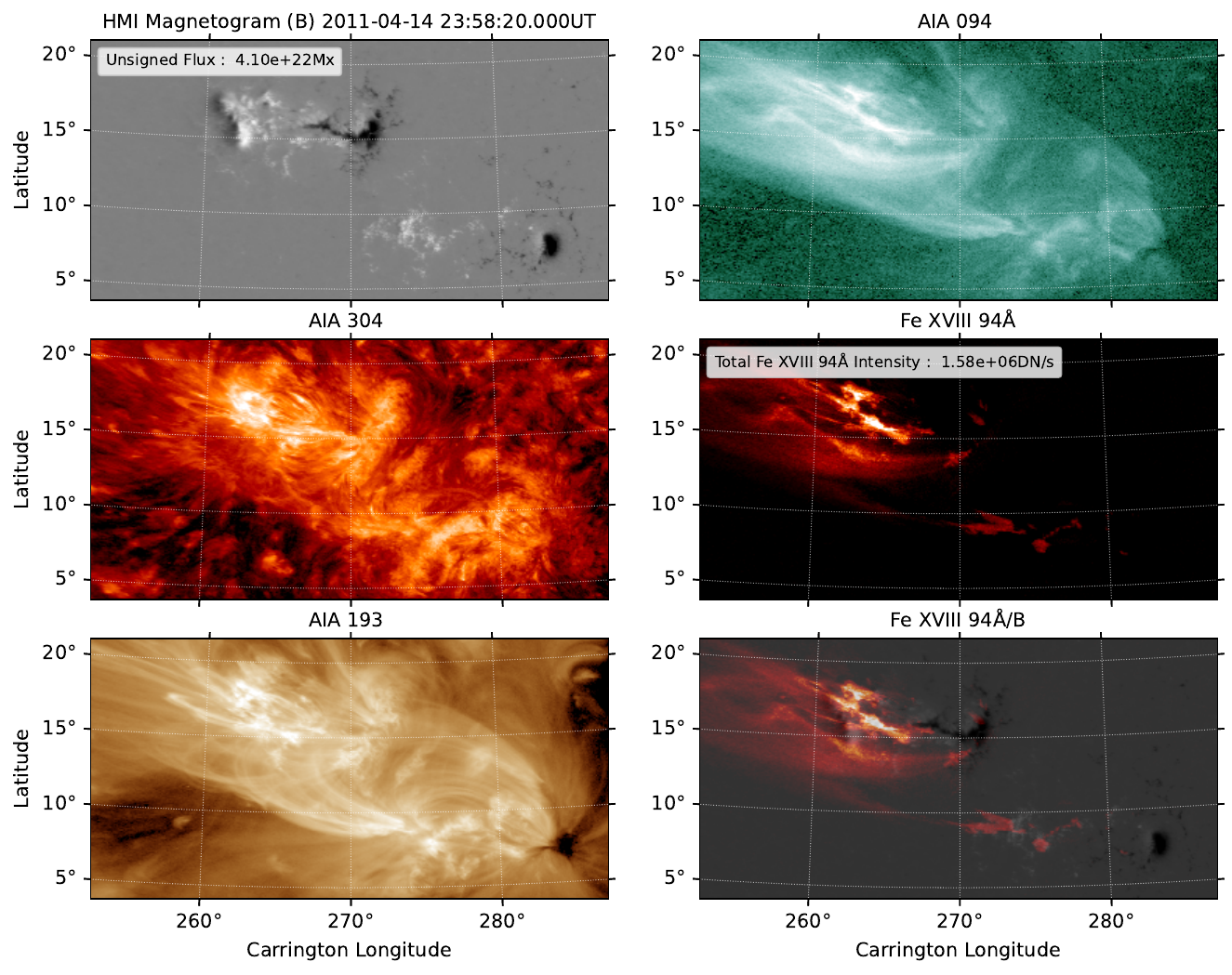}} %
\caption{ Same overview as Figure \ref{fig1} but this time showing AR 11193 observed on 2011 April 14.
\label{fig2}}
\end{figure*}

This null result has stood for over two decades, but it was obtained with the instrumentation available in the 1990s. The vector magnetograms used by \citet{Fisher1998} had limited spatial resolution and temporal coverage. Only a few complexity parameters could be derived. The soft X-ray images provided broad-band emission measures without temperature discrimination. Modern instrumentation enables a more comprehensive test.

We revisit the question using data from the Solar Dynamics Observatory. The Helioseismic and Magnetic Imager (HMI) provides continuous full-disk vector magnetograms at 0.5$''$ resolution \citep{Scherrer2012}. The Space-weather HMI Active Region Patches (SHARPs) data products \citep{Bobra2014} include 16 parameters characterizing magnetic field strength, gradients, vertical currents, current helicity, and photospheric excess magnetic energy—a far more comprehensive characterization of complexity than previously available. The Atmospheric Imaging Assembly (AIA) provides EUV imaging in multiple temperature channels, allowing isolation of hot ($\sim$7 MK) emission from active region cores \citep{Lemen2012}.

We apply partial correlation analysis -- a technique that removes the variance shared with a control variable (here, total unsigned magnetic flux) from both
the predictor and response before computing their correlation -- to isolate the independent contribution of each complexity parameter.
This approach directly addresses the central question: does magnetic complexity contribute to coronal heating beyond what is explained by total magnetic flux?

Note that previous studies have explored the relationship between high temperature emission and both the total unsigned vertical \citep{Fisher1998} and line-of-sight components of the 
magnetic field \citep{Pevtsov2003,Yoshimura2025}. These values can differ for active regions close to the limb or during their emergence phase (when strong horizontal fields impact the 
line-of-sight component). A strong correlation is observed in either case.

Our results, based on a large sample of 137 active regions, confirm the dominance of magnetic flux. Unsigned vertical magnetic flux explains most (72--82\%) of the variance in AIA 304~\AA, 193~\AA, 94~\AA, and Fe~\textsc{xviii} intensity, with a power-law relationship ($I \propto \Phi_M^{0.61-1.19}$). After controlling for flux, most complexity parameters show weak and statistically insignificant partial correlations. One parameter—total photospheric magnetic free energy density—shows potentially significant results. A positive partial correlation with the highest temperature Fe~\textsc{xviii} intensity, and a contrasting 
negative, or non-existent, partial correlation with the lower temperature emission. We discuss possible interpretations of this unexpected result and its implications for coronal heating models.

\section{Data Reduction}

We downloaded 18 monthly batches of SHARPs data spanning the time period from 2010 June 1 to 2011 December 1. This interval covers the early rise phase of Solar
Cycle 24 and avoids long-term instrument calibration drifts, particularly in the AIA EUV channels, that could introduce systematic trends in the analysis.
For each NOAA active region number in each batch, we selected the SHARPs
dataset corresponding to the time when the AR was closest to the central meridian. We then filtered the data to remove duplicate active regions.
From the time-stamps, we downloaded the closest AIA 304~\AA, 193~\AA, and 94~\AA\
images. The emission in these three filters comes from plasma formed in the temperature range of $\sim$0.09--7.1~MK. Correlation analysis of these filters should 
adequately cover the expected temperature range of most active regions. 

This research used version 7.1.0. \citep{sunpy2025} of the SunPy open source software package \citep{sunpy2020}, and
version 0.9.0 of the drms open source software package \citep{Glogowski2019}.
We used the SunPy maps software and DRMS client to download the data from the Joint Science Operations Center (JSOC)  at Stanford University, and reprojected the AIA maps to 
the SHARPs HMI world coordinate system. The AIA data are level 1, so we processed them by updating the pointing, normalizing for exposure time, and correcting
for the time-dependent degradation of the calibration.

From this procedure, we obtained a large sample of active regions with NOAA numbers ranging from 11075 to 11363. We use the 16 
magnetic parameters provided in the SHARPs data headers in our analysis although we recalculate 4 of them.
The magnetic parameters fall into several physical categories: (1) total unsigned flux
($\Phi_M$), which measures the overall magnetic content of the active region; (2) field gradients
($|\nabla B_\mathrm{tot}|$, $|\nabla B_z|$, $|\nabla B_h|$), which quantify how rapidly the
field strength changes spatially and indicate the presence of strong polarity inversion lines;
(3) vertical current density ($J_z$) and related quantities ($J_{z_\mathrm{total}}$,
$J_{z_\mathrm{sum}}$, $\alpha_\mathrm{total}$), which characterize electric currents flowing
along the field; (4) current helicity ($H_c$, $H_{c_\mathrm{total}}$, $H_{c_\mathrm{abs}}$),
which measures the degree of twist in the magnetic field; (5) shear angles ($\gamma$, $\Gamma$,
$A_\mathrm{shear>45^\circ}$), which quantify the angular departure of the observed field from a
potential (current-free) configuration; and (6) photospheric excess magnetic energy density
($\overline{\rho}$, $\rho_\mathrm{tot}$), which estimates the free magnetic energy available for
release. Most (12/16) of the parameter definitions and computation procedures follow \cite{Bobra2014},
and we also adopt their definition of what defines an AR in the magnetic field data.

Several of the SHARP parameters are dependent on the choice of a reference field – namely, those relating to free-magnetic energy ($\bar{\rho}$ and $\rho_\mathrm{tot}$) and magnetic shear ($\bar{\Gamma}$ and $A_\mathrm{shear>45^\circ}$). The SHARP pipeline \citep{Bobra2014} utilizes a Green’s function based representation of the potential field, which is calculated in the local Cartesian approximation from the SHARP CEA-projected radial field maps. 
While suitable for automated analysis of the SHARPs data, this approximation introduces errors in the inferred reference field for large active regions where the Sun's curvature must be taken into account. Additionally, the use of an equal-area projection can lead to distortions in the photospheric field distribution near the edges of large SHARPs FOVs \citep{Calabretta2002}, further affecting the potential field calculation. To address these issues, we recalculated the potential field for each SHARP in our dataset using a spherical harmonic expansion of the full-disk HMI vector magnetograms, as described in \citet{Schuck2022}.
We note that this approach should provide a more accurate estimate of the potential field than the SHARP pipeline, as the proper treatment of the Sun's curvature
eliminates the systematic errors described above.

\begin{figure*}[ht]
  \centering
  \includegraphics[height=0.55\textheight,width=\textwidth]{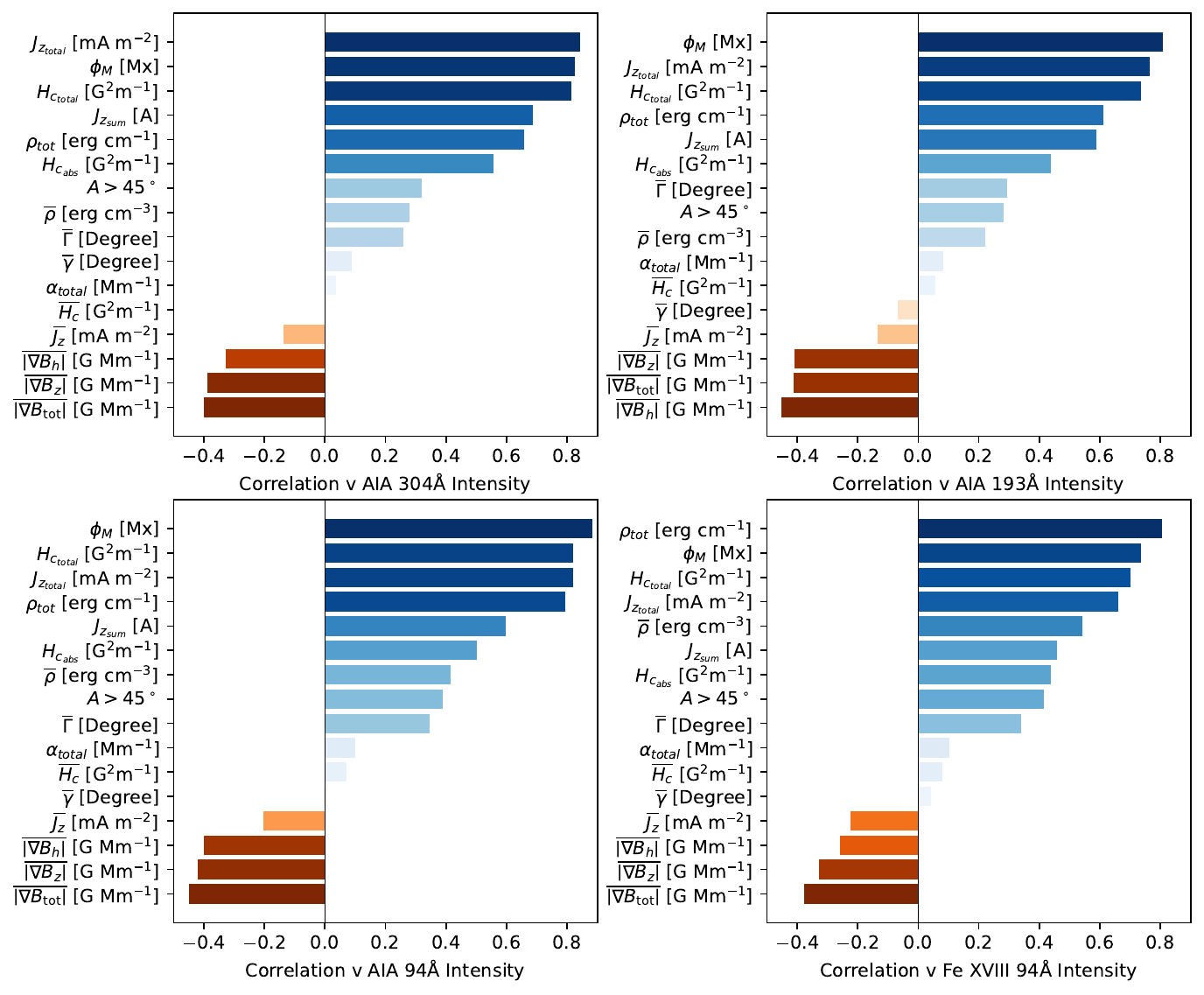}
\caption{ Horizontal bar plots showing the linear Pearson correlation ranking order between each SHARPs magnetic parameter and each AIA filter intensity (including the intensity
for the isolated Fe~\textsc{xviii} 93.926~\AA\ emission). 
Note that the values of 
$\bar{\rho}$, $\rho_\mathrm{tot}$, $\bar{\Gamma}$, and $A_\mathrm{shear>45^\circ}$ are recalculated as discussed in the text.
\label{fig3}}
\end{figure*}

\begin{figure*}[ht]
  \centering
  \includegraphics[height=0.55\textheight,width=\textwidth]{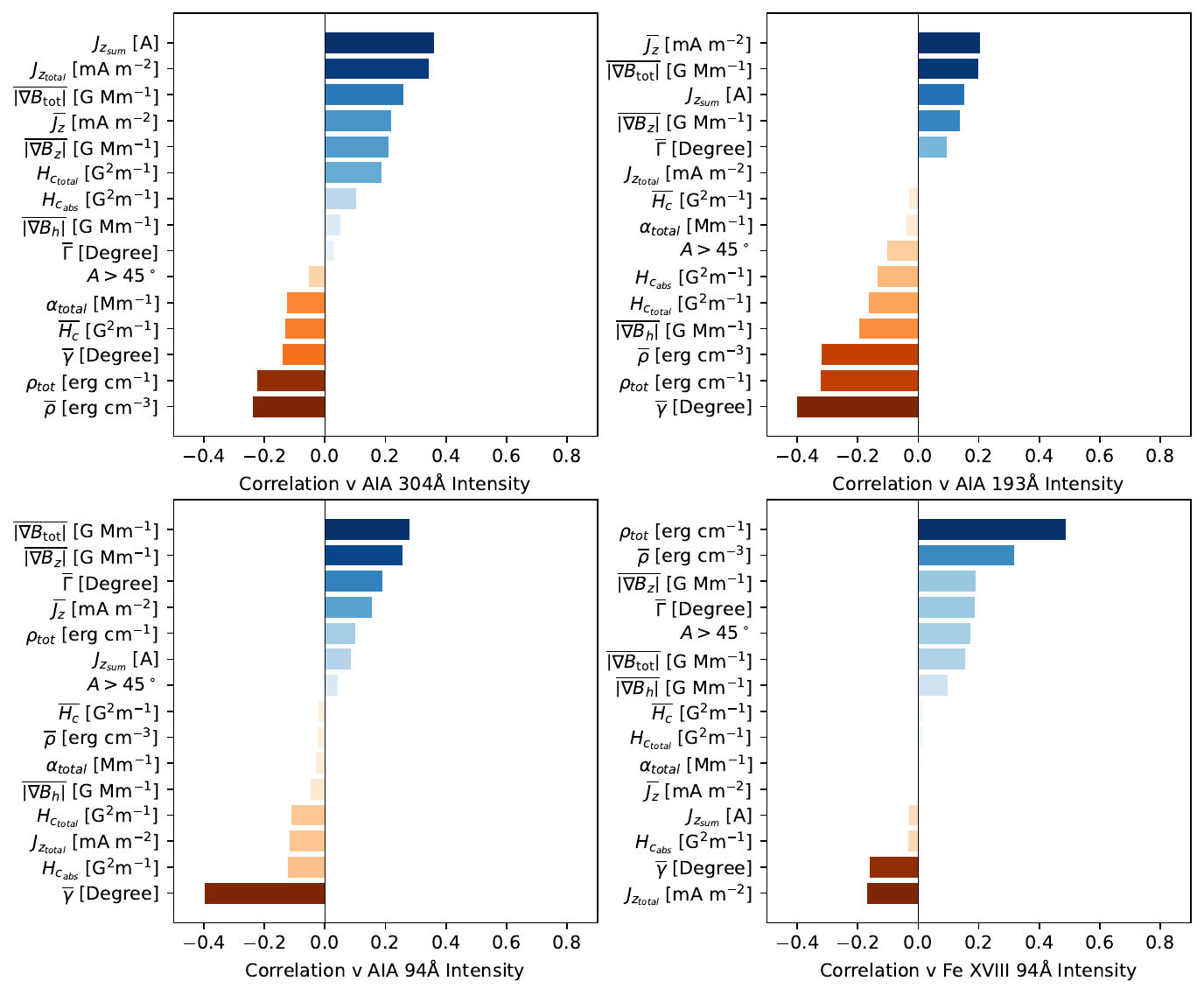}
\caption{  Horizontal bar plots showing the linear Pearson partial correlation ranking order between each SHARPs magnetic parameter and each AIA filter intensity after controlling for total magnetic flux ($\Phi_M$). The x-range of the plots is the same as Figure \ref{fig3} for easier comparison.
Note that the values of 
$\bar{\rho}$, $\rho_\mathrm{tot}$, $\bar{\Gamma}$, and $A_\mathrm{shear>45^\circ}$ are recalculated as discussed in the text.
\label{fig4}}
\end{figure*}

For AIA, we extracted the total intensity in the field-of-view (FOV) for all three filters. Although specialized masks are applied to the SHARP data, for example to
set thresholds on the magnetic field strength and coherence, or assign confidence in the correction for the 180$^{\circ}$ ambiguity, we included all AIA intensities above a 2 DN s$^{-1}$ threshold in our EUV intensity calculation. This is because bright coronal structures cross between magnetic polarities 
and we do not wish to exclude strong emission above non-existent magnetic field patches.

The highest temperature AIA 94~\AA\ filter contains contributions from lower temperature emission characteristic of spectral lines formed closer to 1~MK. 
Any correlations found could potentially be indicative of a relationship with this cooler temperature plasma. We therefore isolate the Fe~\textsc{xviii} 93.926~\AA\
emission using the technique outlined by \cite{Warren2012}. They estimated the contaminating component using a polynomial fit to a mixture of 
the AIA 171~\AA\ and 193~\AA\ images (see equation A1 in the appendix of their paper). By subtracting the modeled cooler emission from the 94~\AA\ image the 
Fe~\textsc{xviii} 93.926~\AA\ is isolated. The method has been cross-checked independently by comparison to SOHO/SUMER observations of Fe~\textsc{xviii} 974.86~\AA\
\citep{Teriaca2012}.

We show overviews of our analysis datasets in Figures \ref{fig1} and \ref{fig2}. These figures show examples of the isolated 
Fe~\textsc{xviii} 93.926~\AA\ emission compared to the AIA 94~\AA\ images, and also the accuracy of the co-alignment between AIA and HMI.
At this stage we also used these summary images to visually assess the entire list of ARs to ensure that no corrupted data, or poorly coaligned
HMI/AIA observations were included in the final analysis. 
During our visual inspection we also excluded cases with only scattered, noisy Fe~\textsc{xviii} 93.926~\AA\ emission.
Sometimes an AR that is clearly visible in the magnetic flux data is not associated with
any coherent emission in Fe~\textsc{xviii} 93.926~\AA. This might be because they were captured in the late decay phase. Sometimes the SHARPs maps
in these cases are much larger than the active region since it is decaying and magnetic concentrations have dispersed to a wider area.
This can potentially lead to an overestimate of the AR high temperature emission since noise associated with quiet surrounding areas is included.
In addition, for small or emerging bipoles, the active area can sometimes be underestimated by the SHARPs defined AR boundary. This could be because
the AR area is growing rapidly, but can also be ambiguous because it is not always clear whether surrounding features should be counted as contributing
to the magnetic flux. In such cases, the total magnetic flux in the SHARPs header can differ from that obtained using a visual inspection. We excluded 
cases where this difference exceeded 10\%.
Our original sample contained 202 ARs, but was ultimately reduced to 137 after inspection.

Table \ref{table1} presents values of all the parameters we study for the first 10 ARs in our sample. The table includes the
NOAA and HARPs numbers for each AR for cross-identification, together
with the date and time of the SHARPs data when the AR was closest to disk center. We provide the AIA 304~\AA, 193~\AA, 94~\AA, and 
Fe~\textsc{xviii} 93.926~\AA\ intensities for each AR, and all 16 SHARPs magnetic parameters. 
The table is an extract from a machine readable version
for all 137 ARs in our study and is available in the electronic version of the manuscript. 

\section{Methods and Results}
We performed three complementary analyses on the complete dataset of 137 ARs:
\begin{enumerate}
    \item \textbf{Pearson and Spearman correlations} between the observed intensities ($I_{304}$, $I_{193}$, $I_{94}$, $I_{\text{Fe}~\textsc{xviii}}$) and each magnetic parameter in Table \ref{table1}.
    \item \textbf{Logarithmic regression} of the observed intensities on $\log(\Phi_M)$ to characterize the power-law relationships.
    \item \textbf{Partial correlations} between the observed intensities and each magnetic parameter, controlling for $\Phi_M$, to identify any independent contributions.
\end{enumerate}
Our findings are as follows.

\subsection{Correlation Analysis}

Figure \ref{fig3} shows the linear Pearson correlation ranking order for each SHARPs magnetic parameter against each of the observed intensities.
The trends for each observed intensity are very similar, though none of them show exactly the same ranking order. In general,
the unsigned magnetic flux shows the strongest correlation ($r = 0.74-0.88$, $p < 0.001$). 
Several other extensive quantities (total current, total helicity) show similarly high correlations - actually the total current is the strongest
for $I_{304}$ - but all of these scale with active region size. Measures of the gradient of the horizontal, vertical, and total field show 
notable (weak to moderate) negative correlations, suggesting that magnetic complexity and rapid field changes are not closely related to the production of high 
temperature emission. While the ranking order is again not exactly the same, the results using the Spearman correlation 
are broadly similar.

\begin{deluxetable}{lcccccc}
\tabletypesize{\small}
\tablecaption{AR parameters}
\tablehead{
\colhead{Date \& Time (UT)} &
\colhead{NOAA} &
\colhead{HARP} &
\colhead{$I_{304}$} &
\colhead{$I_{193}$} &
\colhead{$I_{94}$} &
\colhead{$I_{\text{Fe}~\textsc{xviii}}$}
}
\startdata
2010/06/01 00:00:00  &  11075  &     40  &    3.67E+06  &    2.68E+07  &    1.42E+05 &     3.94E+04   \\
2010/06/01 09:12:00  &  11076  &     43  &    5.87E+06  &    3.22E+07  &    1.96E+05 &     8.43E+03   \\
2010/06/20 10:00:00  &  11082  &     57  &    1.73E+07  &    9.67E+07  &    5.30E+05 &     8.51E+04   \\
2010/07/15 06:24:00  &  11087  &     86  &    4.14E+07  &    2.88E+08  &    1.69E+06 &     3.01E+05   \\
2010/07/25 07:48:00  &  11089  &     92  &    3.61E+07  &    2.50E+08  &    1.62E+06 &     3.10E+05   \\
2010/08/01 00:00:00  &  11092  &    104  &    2.49E+07  &    1.53E+08  &    1.01E+06 &     6.46E+04   \\
2010/08/10 08:36:00  &  11093  &    115  &    3.16E+07  &    1.66E+08  &    8.90E+05 &     4.01E+04   \\
2010/08/09 06:00:00  &  11095  &    114  &    1.10E+07  &    8.09E+07  &    4.02E+05 &     4.11E+03   \\
2010/08/16 10:24:00  &  11097  &    128  &    8.84E+06  &    6.35E+07  &    3.59E+05 &     1.45E+04   \\
2010/08/14 00:00:00  &  11098  &    131  &    5.44E+06  &    2.73E+07  &    1.41E+05 &     7.57E+03   \\
\hline
\multicolumn{1}{c}{$\Phi_M$ \hspace{1.0cm} $\overline{\gamma}$} &
\multicolumn{1}{c}{$\overline{\left|{\nabla B_{\rm tot}}\right|}$} &
\multicolumn{1}{c}{$\overline{\left|{\nabla B_z}\right|}$} &
\multicolumn{1}{c}{$\overline{\left|{\nabla B_h}\right|}$} &
\multicolumn{1}{c}{$\overline{J_z}$} &
\multicolumn{1}{c}{$J_{z_\mathrm{total}}$} & 
\multicolumn{1}{c}{$\alpha_\mathrm{total}$} \\
\hline
  6.18E+20  \hspace{0.1cm}    2.67E+01  &    1.23E+02  &    1.24E+02  &    4.25E+01  &    1.05E+00  &    9.81E+11  &    2.28E-02   \\
  1.57E+21  \hspace{0.1cm}    5.05E+01  &    1.39E+02  &    1.33E+02  &    8.68E+01  &    5.03E-01  &    3.48E+12  &   -3.92E-03   \\
  4.59E+21  \hspace{0.1cm}    4.57E+01  &    1.25E+02  &    1.22E+02  &    7.36E+01  &    7.54E-01  &    8.33E+12  &    1.99E-02   \\
  9.78E+21  \hspace{0.1cm}    4.03E+01  &    1.16E+02  &    1.17E+02  &    5.74E+01  &    7.06E-01  &    1.85E+13  &   -2.49E-02   \\
  1.79E+22  \hspace{0.1cm}    4.37E+01  &    1.19E+02  &    1.19E+02  &    5.90E+01  &   -1.05E-01  &    3.01E+13  &   -4.02E-03   \\
  1.81E+22  \hspace{0.1cm}    4.70E+01  &    8.68E+01  &    8.88E+01  &    4.42E+01  &    3.31E-01  &    2.22E+13  &   -1.36E-02   \\
  6.60E+21  \hspace{0.1cm}    5.58E+01  &    9.17E+01  &    9.22E+01  &    5.73E+01  &    2.62E-01  &    8.92E+12  &   -2.77E-02   \\
  2.71E+21  \hspace{0.1cm}    4.00E+01  &    1.29E+02  &    1.29E+02  &    6.23E+01  &    5.16E-01  &    4.43E+12  &   -1.02E-02   \\
  2.20E+21  \hspace{0.1cm}    3.10E+01  &    1.26E+02  &    1.23E+02  &    5.71E+01  &    3.22E-01  &    3.66E+12  &    5.41E-03   \\
  6.76E+20  \hspace{0.1cm}    4.12E+01  &    1.79E+02  &    1.89E+02  &    9.38E+01  &    1.28E+00  &    1.42E+12  &   -1.33E-02   \\
\hline
\multicolumn{1}{c}{$H_{c_\mathrm{total}}$ \hspace{0.6cm} $H_{c_\mathrm{abs}}$} &
\multicolumn{1}{c}{$J_{z_\mathrm{sum}}$} &
\multicolumn{1}{c}{$\overline{\rho}$} &
\multicolumn{1}{c}{$\rho_\mathrm{tot}$} &
\multicolumn{1}{c}{$\overline{\Gamma}$} &
\multicolumn{1}{c}{$A_\mathrm{shear}>45^\circ$} & 
\multicolumn{1}{c}{$\overline{H_c}$} \\
\hline
  4.08E+01  \hspace{0.1cm}    5.15E+00  &    6.14E+11  &    2.17E+03  &    3.10E+21  &    2.49E+01  &    9.36E+00  &    4.78E-03   \\
  1.71E+02  \hspace{0.1cm}    3.41E+00  &    7.77E+11  &    1.86E+03  &    7.10E+21  &    1.67E+01  &    3.24E+00  &   -1.19E-03   \\
  4.29E+02  \hspace{0.1cm}    5.58E+01  &    3.12E+12  &    3.03E+03  &    3.03E+22  &    2.23E+01  &    8.93E+00  &    7.42E-03   \\
  7.81E+02  \hspace{0.1cm}    1.17E+02  &    6.82E+12  &    1.43E+03  &    3.34E+22  &    1.75E+01  &    4.79E+00  &   -6.60E-03   \\
  1.39E+03  \hspace{0.1cm}    4.10E+01  &    5.88E+12  &    2.27E+03  &    8.93E+22  &    1.94E+01  &    4.98E+00  &   -1.38E-03   \\
  9.66E+02  \hspace{0.1cm}    1.85E+02  &    6.93E+12  &    2.77E+03  &    1.04E+23  &    2.34E+01  &    1.22E+01  &   -6.48E-03   \\
  4.65E+02  \hspace{0.1cm}    1.39E+02  &    1.73E+12  &    1.28E+03  &    1.79E+22  &    1.15E+01  &    7.41E-01  &   -1.32E-02   \\
  1.92E+02  \hspace{0.1cm}    1.42E+01  &    1.29E+12  &    1.74E+03  &    1.05E+22  &    1.71E+01  &    1.90E+00  &   -3.07E-03   \\
  1.54E+02  \hspace{0.1cm}    4.71E+00  &    6.83E+11  &    1.47E+03  &    7.56E+21  &    1.84E+01  &    1.24E+00  &    1.20E-03   \\
  7.87E+01  \hspace{0.1cm}    4.75E+00  &    6.96E+11  &    1.39E+03  &    1.87E+21  &    1.53E+01  &    6.22E+00  &   -4.70E-03   \\
\enddata
\tablenotetext{}{Column (1): Date and Time of the SHARPs data when the AR is closest to disk center.
Column (2): NOAA AR number.
Column (3): Corresponding HARPs number.
Column (4): AIA 304\,\AA\ total intensity of the SHARPs FOV (DN s$^{-1}$).
Column (5): AIA 193\,\AA\ total intensity of the SHARPs FOV (DN s$^{-1}$).
Column (6): AIA 94\,\AA\ total intensity of the SHARPs FOV (DN s$^{-1}$).
Column (7): Fe~\textsc{xviii} 93.926\,\AA\ total intensity of the SHARPs FOV (DN s$^{-1}$).
Column (8): Total unsigned magnetic flux (Mx).
Column (9): Mean magnetic field angle relative to radial (Degrees).
Column (10): Mean horizontal gradient of the total field (G Mm$^{-1}$).
Column (11): Mean horizontal gradient of the vertical field (G Mm$^{-1}$).
Column (12): Mean horizontal gradient of the horizontal field (G Mm$^{-1}$).
Column (13): Vertical current density (mA m$^{-2}$).
Column (14): Total unsigned vertical current (A).
Column (15): Magnetic twist (M m$^{-1}$).
Column (16): Total unsigned vertical field contribution to current helicity (G$^2$ m$^{-1}$).
Column (17): Absolute value of net vertical field contribution to current helicity (G$^2$ m$^{-1}$).
Column (18): Total modulus of the net current per polarity (A).
Column (19): Mean photospheric excess magnetic energy density (erg cm$^{-3}$).
Column (20): Total photospheric magnetic free energy density (erg cm$^{-1}$).
Column (21): Magnetic shear angle (Degrees).
Column (22): Fractional area with high shear.
Column (23): Vertical field contribution to current helicity (G$^2$ m$^{-1}$).
Note that the values of 
$\bar{\rho}$, $\rho_\mathrm{tot}$, $\bar{\Gamma}$, and $A_\mathrm{shear>45^\circ}$ are recalculated as discussed in the text.
Only a portion of this table (10 ARs) is shown here to demonstrate its form and content. A machine-readable version of the full table including all 137 ARs is available.}
\label{table1}
\end{deluxetable}

\begin{figure*}[ht]
    \centering
    \includegraphics[width=\textwidth]{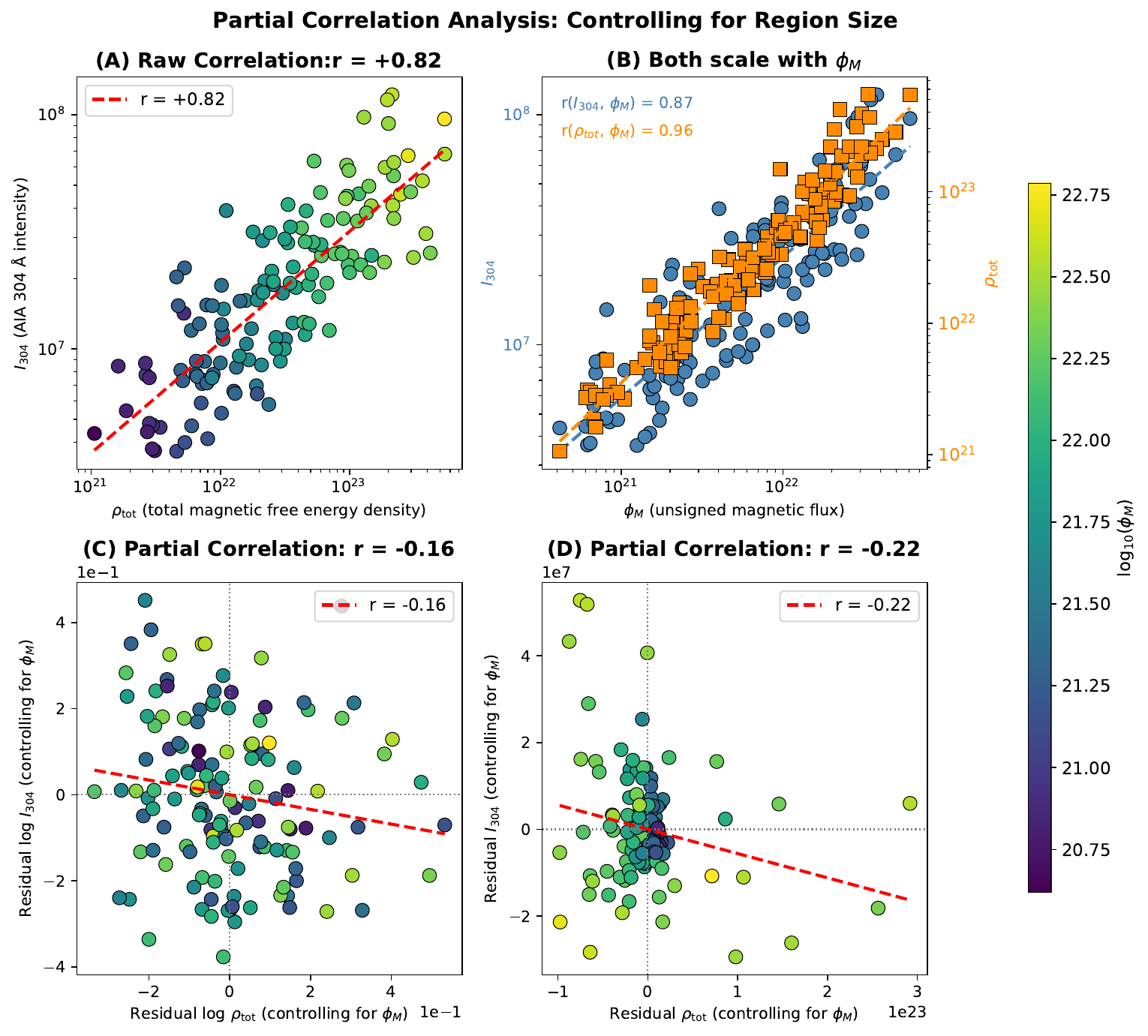}
    \caption{Illustration of the partial correlation analysis for $\rho_\mathrm{tot}$ vs $I_{304}$. (A)~Raw correlation shows a strong positive relationship. (B)~Both variables scale with unsigned flux $\Phi_M$, suggesting the correlation in (A) is driven by region size. (C)~Partial correlation between logarithmic residuals shows no relationship. (D)~Partial correlation between linear residuals reveals a weak \emph{negative} relationship: regions with excess magnetic energy (for their size) have lower 304~\AA\ emission. Note that this analysis uses the recalculated value of $\rho_\mathrm{tot}$.}
    \label{fig5}
\end{figure*} 

\begin{figure*}[ht]
    \centering
    \includegraphics[width=\textwidth]{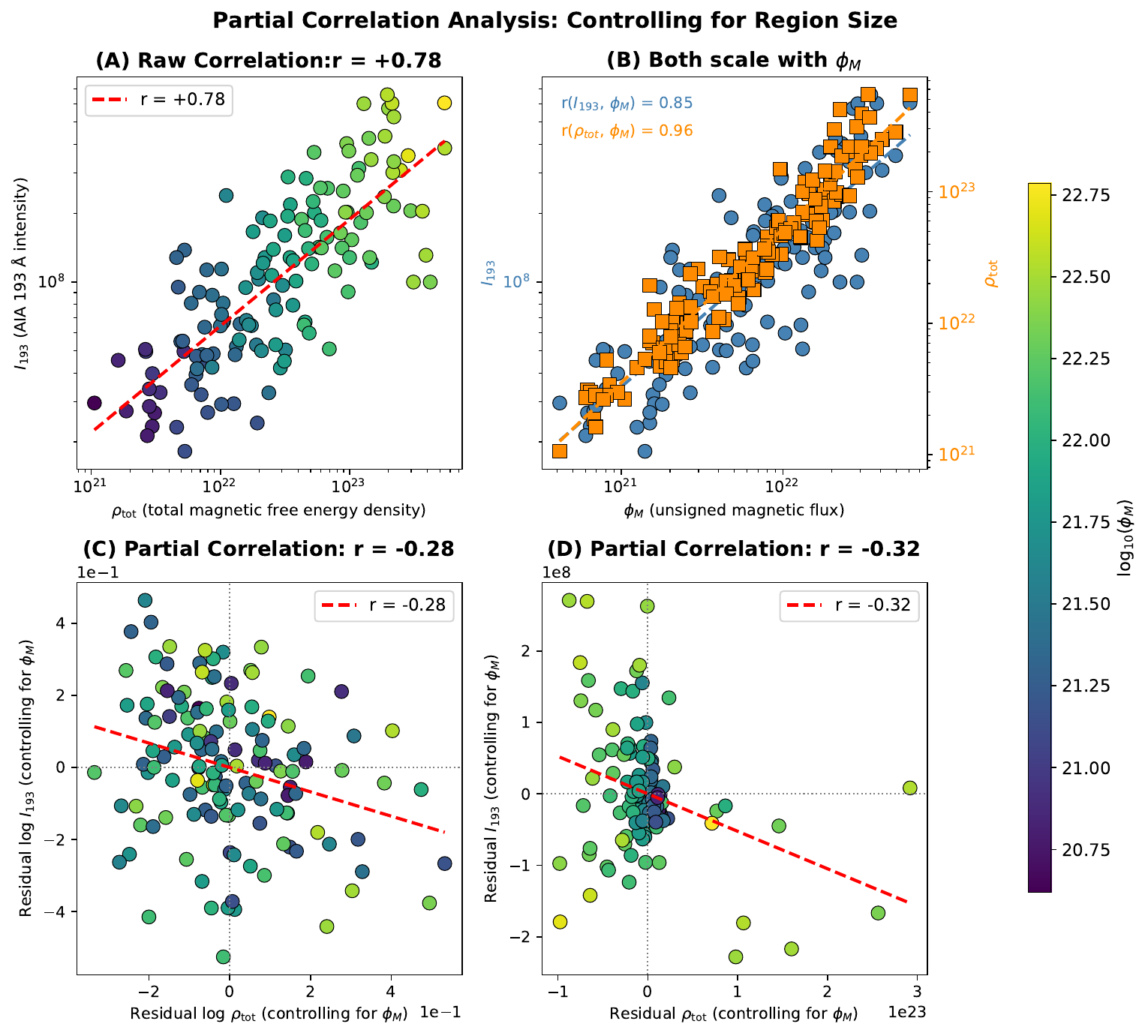}
    \caption{Illustration of the partial correlation analysis for $\rho_\mathrm{tot}$ vs $I_{193}$. (A)~Raw correlation shows a strong positive relationship. (B)~Both variables scale with unsigned flux $\Phi_M$, suggesting the correlation in (A) is driven by region size. (C)~Partial correlation between logarithmic residuals shows no relationship. (D)~Partial correlation between linear residuals reveals a weak \emph{negative} relationship: regions with excess magnetic energy (for their size) have lower 193~\AA\ emission. Note that this analysis uses the recalculated value of $\rho_\mathrm{tot}$.}
    \label{fig6}
\end{figure*} 

\begin{figure*}[ht]
    \centering
    \includegraphics[width=\textwidth]{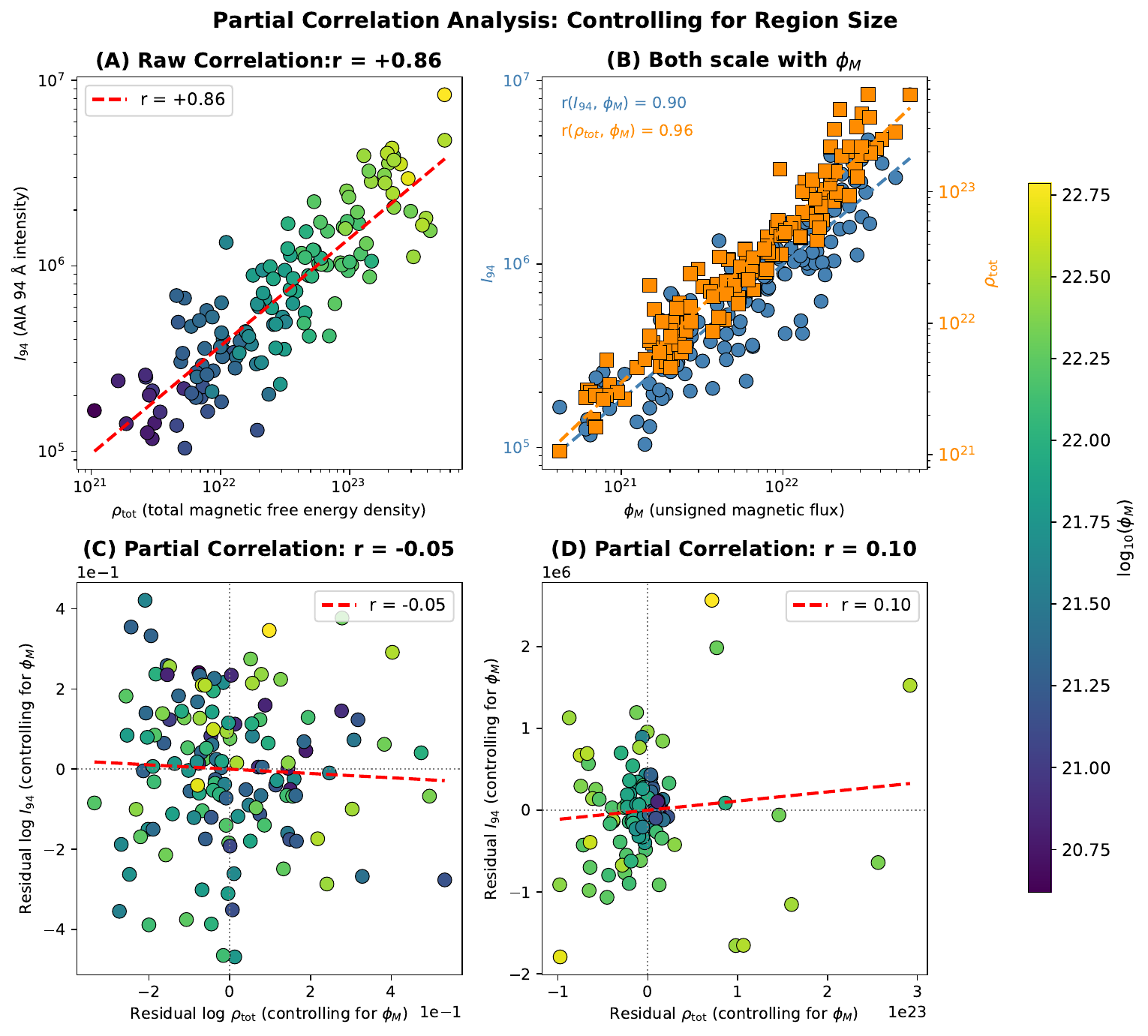}
    \caption{Illustration of the partial correlation analysis for $\rho_\mathrm{tot}$ vs $I_{94}$. (A)~Raw correlation shows a strong positive relationship. (B)~Both variables scale with unsigned flux $\Phi_M$, suggesting the correlation in (A) is driven by region size. (C)~Partial correlation between logarithmic residuals shows no relationship. (D)~Partial correlation between linear residuals also shows no relationship in this case. Note that this analysis uses the recalculated value of $\rho_\mathrm{tot}$.}
    \label{fig7}
\end{figure*} 

\begin{figure*}[ht]
    \centering
    \includegraphics[width=\textwidth]{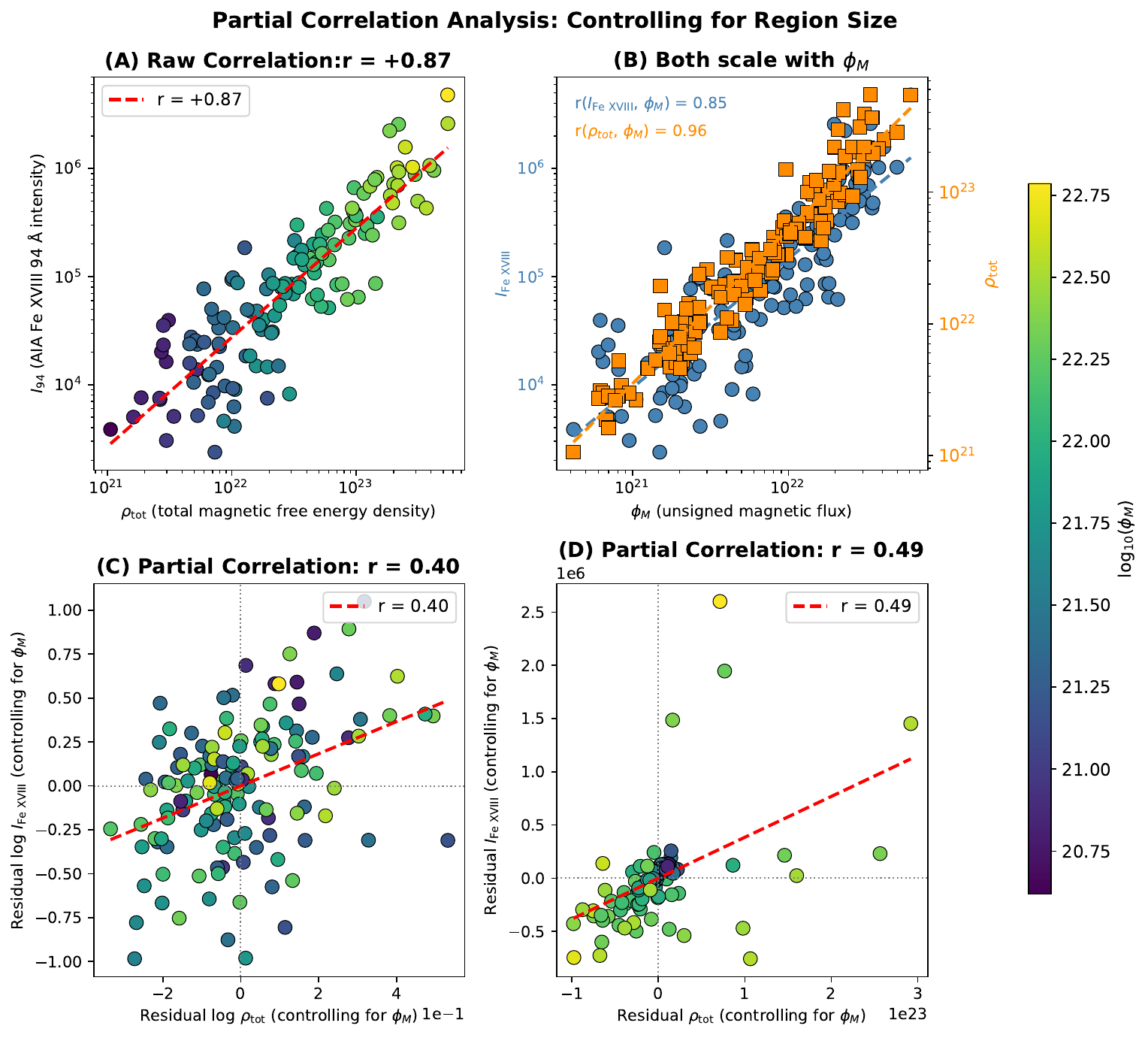}
    \caption{Illustration of the partial correlation analysis for $\rho_\mathrm{tot}$ vs $I_{\text{Fe}~\textsc{xviii}}$. (A)~Raw correlation shows a strong positive relationship. (B)~Both variables scale with unsigned flux $\Phi_M$, suggesting the correlation in (A) is driven by region size. (C)~Partial correlation between logarithmic residuals reveals a \emph{positive} relationship. (D)~Partial correlation between linear residuals also shows a \emph{positive} relationship: regions with excess magnetic energy (for their size) have higher Fe~\textsc{xviii}~\AA\ emission. Note that this analysis uses the recalculated value of $\rho_\mathrm{tot}$.}
    \label{fig8}
\end{figure*} 

\subsection{Power-Law Relationships}

To characterize the power-law relationships between the observed intensities and the unsigned flux, $\Phi_M$, we performed linear regression
analysis on $\log(\Phi_M)$. The regression analysis yielded:
\begin{equation}
\begin{aligned}
    \log(I_{304})               &= 0.61 \log(\Phi_M) - 6.08  \\
    \log(I_{193})               &= 0.62 \log(\Phi_M) - 5.42  \\
    \log(I_{94})                &= 0.74 \log(\Phi_M) - 10.23  \\
    \log(I_{\text{Fe}~\textsc{xviii}}) &= 1.19 \log(\Phi_M) - 21.05 
\end{aligned}
\end{equation}
with $R^2 = 0.75, 0.72, 0.82,$ and $0.72$, respectively, and $p < 0.001$ in all cases. These
correspond to power-laws:
\begin{equation}
\begin{aligned}
    I_{304}               &\propto \Phi_M^{0.61} \\
    I_{193}               &\propto \Phi_M^{0.62} \\
    I_{94}                &\propto \Phi_M^{0.74} \\
    I_{\text{Fe}~\textsc{xviii}} &\propto \Phi_M^{1.19}
\end{aligned}
\end{equation}
with close-to linear exponents of different values for each intensity.
These results show that the power-law exponent varies considerably (0.58) with temperature.

Note that \cite{Fisher1998} found $L_X \propto \Phi_M^{1.19}$. The closest comparison between their soft X-ray results and our measurements
is for the highest temperature emission ($I_{\text{Fe}~\textsc{xviii}}$).
Although the exponent for the AIA 94~\AA\ filter is different (0.74), the results match exactly after isolating 
the Fe~\textsc{xviii} 93.926~\AA\ line (1.19).

The sub-linear exponents for $I_{304}$, $I_{193}$, and $I_{94}$ suggest that larger active regions produce proportionally less hot plasma per unit magnetic flux, possibly indicating saturation of the heating mechanism or geometric effects.
The power-law for $I_{Fe~\textsc{xviii}}$ is almost proportional to the unsigned flux whereas the exponents for the other intensities are lower. This suggests a 
significant difference in the accumulation of high temperature material, compared to lower temperatures, in larger active regions. It also suggests a proportionally
increasing ratio of high to low temperature material in growing active regions. This is interesting in the context of the slope of the emission measure distribution
and the relationship with heating frequency \citep{Ugarte2012}.

\subsection{Partial Correlation Analysis}

Several magnetic parameters correlate strongly with the observed intensities, but this may simply reflect the fact that both quantities scale with active region size (i.e., $\Phi_M$). To identify parameters that influence emission \emph{independently} of region size, we computed partial correlations controlling for $\Phi_M$.

The partial correlation between observed intensity $I_{i}$, and a magnetic parameter $X$, controlling for $\Phi_M$, is computed as follows:
\begin{enumerate}
    \item Regress $I_{i}$ on $\Phi_M$ and compute residuals: $\epsilon_{I} = I_{i} - \hat{I}_{i}(\Phi_M)$
    \item Regress $X$ on $\Phi_M$ and compute residuals: $\epsilon_{X} = X - \hat{X}(\Phi_M)$
    \item Compute the Pearson correlation between the residuals $\epsilon_{I}$ and $\epsilon_{X}$
\end{enumerate}

This procedure removes the variance in both variables that is attributable to $\Phi_M$, and therefore reveals any residual relationship. 

Figure \ref{fig4} shows the linear Pearson partial correlation ranking order for each SHARPs magnetic parameter against each of the observed intensities
after controlling for $\Phi_M$. The trends for each observed intensity are less coherent than those seen in Figure \ref{fig3} and the correlations (if any) are
generally weak. Correlations between the gradient of the horizontal, vertical, and total field and the EUV intensities, that were all negative, 
become weaker and are no longer directionally consistent. 
Some measures of the current and field gradient show weak partial correlations of $\sim$0.21--0.36 with AIA 304~\AA. 
$I_{193}$ and $I_{94}$
also show negative partial correlations (-0.4) with mean field angle.
The most significant partial correlation
is between the highest temperature emission ($I_{Fe~\textsc{xviii}}$) and $\rho_\mathrm{tot}$ (0.49).

Figure~\ref{fig5} illustrates the analysis for $I_{304}$ and $\rho_\mathrm{tot}$. The raw correlation between the parameters is strong ($r = 0.82$), but
the even stronger correlation between both parameters and the unsigned flux ($r > 0.87$) suggests that it is driven by active region size. 
Partial
correlation of the residuals shows a very weak \emph{negative} relationship between the power laws in log space, that strengthens ($r = -0.22$)
in linear space. 
This value is robust against a statistical test where we remove the top 10\% of the values of $\rho_\mathrm{tot}$.
Residuals from regressing $I_{304}$ on $\Phi_M$ in linear space represent the emission variation at fixed magnetic flux.

Figures~\ref{fig6}-\ref{fig8} illustrate the analysis for $I_{193}$, $I_{94}$, and $I_{Fe~\textsc{xviii}}$ against $\rho_\mathrm{tot}$. The basic picture is mostly as already described, with
strong raw correlations $r = 0.78-0.87$ driven by even stronger correlations with the unsigned flux $r = 0.85-0.96$ and therefore active region size. The residual partial correlations
with $\rho_\mathrm{tot}$ depend on wavelength. $I_{193}$ shows a similar result to $I_{304}$, with a weak
 \emph{negative} relationship in log space that strengthens ($r = -0.32$) in linear space. 
Conversely, $I_{94}$ shows no relationship with $\rho_\mathrm{tot}$.
These values are also fairly robust against a statistical test where we remove the top 10\% of the values of $\rho_\mathrm{tot}$; changing to $r = -0.18$
and $r = 0.21$ for $I_{193}$ and $I_{94}$, respectively.

The lack of a relationship with $I_{94}$ is perhaps
surprising since, as we noted, the strongest residual partial correlation is found for Fe~\textsc{xviii} 93.926~\AA: in linear space, $\rho_\mathrm{tot}$ shows a moderate ($r = 0.49$) \emph{positive} 
relationship with the Fe~\textsc{xviii} 93.926~\AA\ intensity. It may be that this positive relationship is masked in the AIA 94~\AA\ filter by the negative relationship associated with the EUV emission at lower temperatures,
and only appears when the Fe~\textsc{xviii} 93.926~\AA\ emission is isolated. As with the power-laws, this result also suggests a difference in the accumulation of higher temperature material
compared to lower temperatures. Note that the relationship becomes significantly stronger if we perform the same statistical test as before. Removing the
top 10\% of the values of $\rho_\mathrm{tot}$ increases $r$ to 0.68.

Another possible explanation for these different relationships with $\rho_\mathrm{tot}$ is that additional heating from non-potential fields due to electric currents
raises the plasma temperature to higher than average values that only Fe~\textsc{xviii} 93.926~\AA\ is sensitive to. Some of the plasma is thus heated out of visibility in the 
low temperature channels resulting in the observed apparent suppression of EUV emission by field complexity in those filters. Fe~\textsc{xviii} 93.926~\AA\ then shows the
observed positive correlation. The AIA 94~\AA\ filter shows a null correlation because it samples both the low and high temperature components i.e. plasma heated out of
the lower temperature response gets elevated into the high temperature response within the same filter.

In summary, the positive correlations between the observed intensities
and $\rho_\mathrm{tot}$  become negative or non-existent for $I_{304}$, $I_{193}$, and $I_{94}$ after controlling for magnetic flux. 
Active regions with \emph{higher} photospheric excess magnetic energy show
\emph{lower} EUV emission in these AIA filters. In contrast, once the high temperature Fe~\textsc{xviii} 93.926~\AA\ emission is isolated, it shows \emph{higher} EUV emission in
active regions with \emph{higher} photospheric excess magnetic energy. This suggests that including $\rho_\mathrm{tot}$ in the model may be a better predictor of
high temperature Fe~\textsc{xviii} 93.926~\AA\ emission from magnetic field data.

\section{Discussion and Conclusions}

We have presented a comprehensive correlation analysis between AIA observed intensities and 16 magnetic parameters from the SHARPs database. The AIA intensities cover the transition region (304~\AA) to the corona (193~\AA, 94~\AA) for a collection of 137 ARs spanning $\sim$2 orders of magnitude in vertical magnetic flux (4.2$\times$10$^{20}$ -- 6.1$\times$10$^{22}$~Mx). We isolated the Fe~\textsc{xviii} 93.926~\AA\ component of the 94~\AA\ channel by removing the lower temperature contributions. We also recalculated several SHARPs parameters using improved potential reference fields.

Our analysis shows that the total unsigned magnetic flux is the main driver of AR EUV emission. For example, the unsigned flux explains $\sim$82\% of the variance in AIA 94~\AA\ intensity. This is consistent with previous studies using lower resolution observations from SOHO, Yohkoh, and Hinode \citep{Fisher1998,Pevtsov2003,Fludra2008,Yoshimura2025}.

We find that the EUV emission scales with unsigned flux by power-laws. The high temperature emission (Fe~\textsc{xviii} 93.926~\AA) is closest to linear, suggesting that larger ARs accumulate more higher temperature material.

Partial correlations controlling for total magnetic flux reveal interesting relationships with the total excess magnetic energy density ($\rho_\mathrm{tot}$). The lower temperature emission (AIA 304~\AA\ and AIA 193~\AA) shows a negative correlation, suggesting an inverse scaling with AR complexity.  
One possible interpretation is that high magnetic stress represents stored energy that has not yet been released. Active regions with elevated $\rho_\mathrm{tot}$ may be in a pre-flare state, with energy accumulating in non-potential magnetic configurations rather than being dissipated as coronal heating.
This is consistent with ideas that suggest magnetic complexity impedes ordinary coronal heating, and is removed by flaring activity in the early stages of AR evolution \citep{Reale2019,Testa2023,Brooks2026}.

In contrast, however, the AIA 94~\AA\ emission shows no relationship with $\rho_\mathrm{tot}$, and the correlation changes sign to positive when the highest temperature (Fe~\textsc{xviii} 93.926~\AA) emission is isolated. This suggests 
a significant difference in how high temperature material accumulates in larger active regions compared to lower temperatures, and that the relationship might be masked by lower temperature emission in the 94~\AA\ filter. 
Previous observations have also detected evidence of differences in the structure of the magnetic field associated with low- and high-temperature emission. 
\cite{Brooks2008}, for example, showed that transient heating events in the chromosphere (He~\textsc{ii} 256.35~\AA, so comparable to AIA~304\AA) 
occur preferentially around the polarity inversion line, where the vertical field gradient changes rapidly. In contrast, higher temperature (Fe~\textsc{xvi} 262.94~\AA) loops 
are rooted in locally unipolar regions of unmixed magnetic flux. 

An important caveat is that our analysis compares active-region-averaged measures of magnetic
complexity with spatially integrated EUV emission. While our results suggest that magnetic complexity only
contributes to the formation of the highest temperature emission in active regions,
it is possible that it plays
a significant role at lower temperatures at smaller spatial scales---for example, at individual loop footpoints or
along polarity inversion lines---but that this relationship is diluted or lost when both
quantities are averaged over the full active region. Spatially resolved studies that correlate
local measures of magnetic complexity with local coronal emission would be needed to address
this question.

EUV emission is not the only non-magnetic observable. Future work will consider the role of other spectroscopic measurements (flows and motions, plasma composition). A recent paper \citep{McKevitt2026} has suggested that spectral line broadening at 1.6~MK averaged over whole ARs is also moderately correlated with the total unsigned magnetic flux. Interestingly, considering the above comment on spatially resolved studies, this correlation is not seen in individual loops \citep{Brooks2016}. In any case, at present, the total unsigned flux is the most reliable predictor of high temperature AR emission.

\begin{acknowledgments}
We thank the reviewer for the careful data check that resulted in an improved region
filtering process and strengthened our results.
We also thank Ignacio Ugarte-Urra for helpful discussions on this work.  
The work was supported by the Office of Naval Research 6.1 Program and the 
NASA Hinode program. 
The AIA and HMI data are courtesy of the NASA/SDO and the AIA, EVE, and HMI science teams.
\end{acknowledgments}

\facilities{SDO}

\end{document}